\documentclass{icrc2009}

\usepackage{graphicx}   
\usepackage[caption=false]{caption}    
\usepackage[font=footnotesize]{subfig} 
\usepackage{fixltx2e}
\usepackage{url}

\newcommand{\shorttitle}[1]%
{\markboth{Proceedings of the 31\MakeLowercase{$^{st}$} ICRC, {\L}\'{o}d\'{z} 2009}{#1} }
\newcommand{\etal}{\MakeLowercase{\textit{et al. }}} 

\begin{document} 

\title{First tests and long-term prospects of Geigermode avalanche 
photodiodes as camera sensors for IACTs}

\author{\IEEEauthorblockN{Eckart Lorenz\IEEEauthorrefmark{1}\IEEEauthorrefmark{2}, Razmick
Mirzoyan\IEEEauthorrefmark{1}, Hiroko Miyamoto\IEEEauthorrefmark{1}, Nepomuk
Otte\IEEEauthorrefmark{3}, and Dieter Renker\IEEEauthorrefmark{4}} \\
\IEEEauthorblockA{\IEEEauthorrefmark{1} Max Planck Institute for Physics,
F\"{o}hringer Ring 6, 80805 Munich, Germany}
\IEEEauthorblockA{\IEEEauthorrefmark{2} ETH Zurich, CH-8093 Z\"{u}rich, Switzerland}
\IEEEauthorblockA{\IEEEauthorrefmark{3} Dep. of Physics, UC Santa Cruz, 1156
High St., Santa Cruz, California, CA 95064, USA}
\IEEEauthorblockA{\IEEEauthorrefmark{4} Paul Scherrer Institute, CH-5232 Villigen,
Switzerland}}


\shorttitle{Eckart Lorenz \etal Photosensors for IACTs}
\maketitle

\begin{abstract}
Geigermode avalanche photodiodes (G-APD) are novel photodetectors, 
which can detect single photons. This diode might become an alternative to 
photomultipliers (PMT) in next generation IACTS. Prospects, limitations and 
development directions will be discussed. Results from first tests will be 
reported. 
\end{abstract}


\section{Introduction}

Very high energy (VHE) gamma-ray ($\gamma )$ astronomy, a section of high 
energy astroparticle physics, has begun in 1989 and is currently a very 
successful field of fundamental research. Most discoveries have been 
achieved by experiments using so-called imaging atmospheric Cherenkov 
telescopes (IACT). These telescopes detect the very faint but very fast 
Cherenkov light flashes from air showers initiated by cosmic particles 
hitting the upper atmosphere. IACTs require high detection efficiency for 
Cherenkov photon flashes and fast time resolution to filter out the few nsec 
lasting flashes against the steady night sky light background ($\ge $ 
2.10$^{12}$ photons/m$^{2}$sec sterad, between 300 and 600 nm). Large 
mirrors project the Cherenkov light onto a fine pixelized matrix of ultra 
sensitive photosensors in the focal plane. By recording the shower light one 
obtains a coarse image of the shower. Using a detailed analysis of the 
shower image one can determine the energy of the primary particle, its 
origin in the sky map (only for neutral particles) and discriminate $\gamma 
$ shower candidates against the many orders more frequent hadronic showers. 
Further information on the detection methods can be found for example in 
[Weekes]. 

In order to make further progress in this research field one has to increase 
the sensitivity of IACTs and to lower the energy threshold by detecting more 
Cherenkov photons/shower. This can be accomplished by either increasing the 
area or/and the photosensor detection efficiency. Beyond 25-30 m mirror 
diameter optical errors as well as material parameters set practical limits. 
Another option is to improve the photon detection efficiency (PDE) of the 
light sensors. Up to now IACTs use photomultipliers (PMT) with a typical 
peak quantum efficiency (QE) of 25{\%} and a rather narrow sensitivity 
range. Recently, industry has achieved raising the peak QE to 32 
respectively to 42 {\%} for the so-called superbialkali (Sba) and 
ultrabialkali (Uba) photocathodes. PMTs with Sba cathodes will be used for 
the first time in the 17 m {\O} MAGIC II telescope \cite{borla}. 

Since a few years a new silicon semiconductor photosensor has been under 
investigation. It has very high gain based on secondary avalanche 
multiplication and quenching similar to the Geiger counter principle, 
therefore generically named Geigermode avalanche photodiodes (G-APD). G-APDs 
(also called SiPM, MPPC, MPGM-APD, SSPM, MRS-APS, MKPD..) have the potential 
to reach a QE of up to 80{\%} in the spectral range matched to Cherenkov 
light. The PDE is currently substantially lower but it is expected that 
rather soon significant improvements will be achieved thus offering a 
quantum jump in sensitivity for IACTs. In the following we will discuss in 
chapter II the detection principle and in chapter III some very first 
Cherenkov light observations carried out in 2007. In chapter IV we comment 
on some deficiencies and the developments necessary to reach the required 
performance for IACTs. The paper will conclude in chapter V with a short 
outlook.

\section{The G-APD as a detector for single photons.} 

When reverse biasing a semiconductor p-n structure, such as for example a 
high quality silicon diode, the p-n zone is depleted and acts as an isolator 
with only a small leakage current caused by thermally generated 
electron-hole (e-h) pairs. When the field strength across the p-n structure 
is increased to $\approx $ 140 kV/cm the accelerated electrons can produce 
secondary e-h pairs by impact ionization thus starting avalanche 
multiplication. As a result an amplification of the current occurs but 
avalanches develop only from the p towards the n layer, as holes are not 
accelerated high enough to initiate also avalanches. In this state of 
biasing also external photons are able to initiate avalanches by e-h 
generation by absorption. The device acts as a linear mode avalanche 
photodiode with modest gain. The signal is in first order proportional to 
the photon flux. If the field strength across the p-n structure is further 
increased some accelerated holes can also produce e-h pairs by impact 
ionization. This process will start secondary avalanches on both sides of 
the pn structure and in turn a sustained current will flow provided the 
diode is still biased above the so-called breakdown voltage (V$_{bd})$. This 
can eventually lead to the destruction of the diode. By adding a series 
resistor between the bias source and the diode the avalanche current will 
result in a voltage drop, which eventually will quench the multiplication 
process once the voltage across the diode drops below V$_{bd}$- similar to 
the process in Geiger counters. In such a configuration a single electron 
can initiate a large signal of standard amplitude given by the diode 
capacitance and the overvoltage above the breakdown voltage. The gain G is: 

G = (1/q)* C$_{diode}$*(V-V$_{bd})$ (1) 

with q being the elementary charge.

The total number of charge carriers in the avalanche process is therefore 
independent of the number of initial e-h pairs, i.e. such a diode is a 
digital counter. An important feature for using such diodes as photon 
counters is the fact that small volumes of appropriately designed diodes can 
be biased for quite some time well above V$_{bd}$ because the probability of 
a thermally generated e-h pair in the small volume can be quite low. Such 
diodes can act as single photon counters with a very high gain up to a few 
10$^{6}$. It should be noted that after each discharge the diode must be 
charged up again. During the charge-up time, set by the product of the 
capacitance and the quenching resistor, the diode is basically insensitive 
to new photons. Such detectors of very small area were already used in the 
late sixties for single photon detection. Around 1990 the concept was 
modified by some Russian physicists by combining many small diodes, each 
with its own integrated quenching resistor connected to a bus, onto a 
silicon wafer. The basic element (hereafter called cell) of a so-called 
G-APD is a miniature photodiode operated slightly above V$_{bd}$. Fig 1 
shows the schematics of such a G-APD. Due to the standardized signal from 
each cell the G-APD has a very good single electron resolution allowing one 
to resolve up to a few tens of photoelectrons (PE). Due to the internal 
photoeffect a high quantum efficiency (QE) can in principle be achieved, 
i.e. much higher than with PMTs.

 \begin{figure}[!t]
  \centering
  \includegraphics[width=2in]{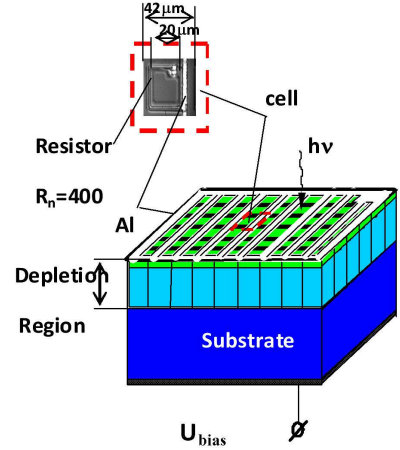}
  \caption{Basic configuration of a Geigermode apd named SiPm (courtesy B.
  Dolgoshein).}
  \label{fig1}
 \end{figure}

On the other hand the small cells must be separated by some inefficient 
material leading to a lower QE when averaged over the entire area. Cells 
have typically a sensitive area up to 100x100 microns and a few micron wide 
dead areas around them. Therefore, at best a QE of 20-70 {\%} can be 
achieved in standard G-APDs. Such QE can hardly be reached in the high gain 
counting mode because not all electrons or holes will initiate an avalanche. 
The probability depends on the diode type and also on the overvoltage above 
breakdown. A typical photon detection efficiency (PDE) of a blue sensitive 
p-on-n G-APD set at $\approx $ 1V overvoltage is about 50 {\%} of the QE but 
can reach $\approx $ 90{\%} at $\ge $ 4 V. Therefore, it is important for 
performance estimates to replace the QE by the PDE, which is both a function 
of the QE and the overvoltage (V-V$_{bd})$. Currently, G-APDs are still in 
an advanced state of development. They have some limitations but quite some 
potential to replace PMTs in IACTs in a few years. As we simplified the 
description of the detection process, we refer the interested reader to 
\cite{renker} referring also the relevant references. 

Advantages of G-APDs for IACT cameras are:

\begin{itemize}
\item A high PDE 
\item Very compact ($<$ 2 mm thickness)
\item High gain
\item Insensitive to magnetic fields
\item Low bias voltage ($<$ 100 V)
\item Very fast time resolution
\item Not damaged when exposed biased to day light
\item Potential for low cost
\end{itemize}
Disadvantages are:

\begin{itemize}
\item PDE significantly lower than the QE
\item PDE depends on overvoltage
\item Performance temperature dependent
\item PDE sensitive to small voltage variation
\item UV sensitivity needs improvement
\item Optical cross talk
\item Noise rate ($\ge $100 kHz/mm$^{2 }$at room temperature)
\item Small size (currently $\le $ 0.3 cm$^{2})$
\item Dynamic range limited
\item Currently too expensive
\end{itemize}
Most of these deficiencies can either be corrected or require a stabilized 
bias voltage or need constructive changes to make the G-APD a superior 
lightsensor for IACT cameras, see chapter 4.

 \begin{figure}[!t]
\centering
  \includegraphics[width=2.1in]{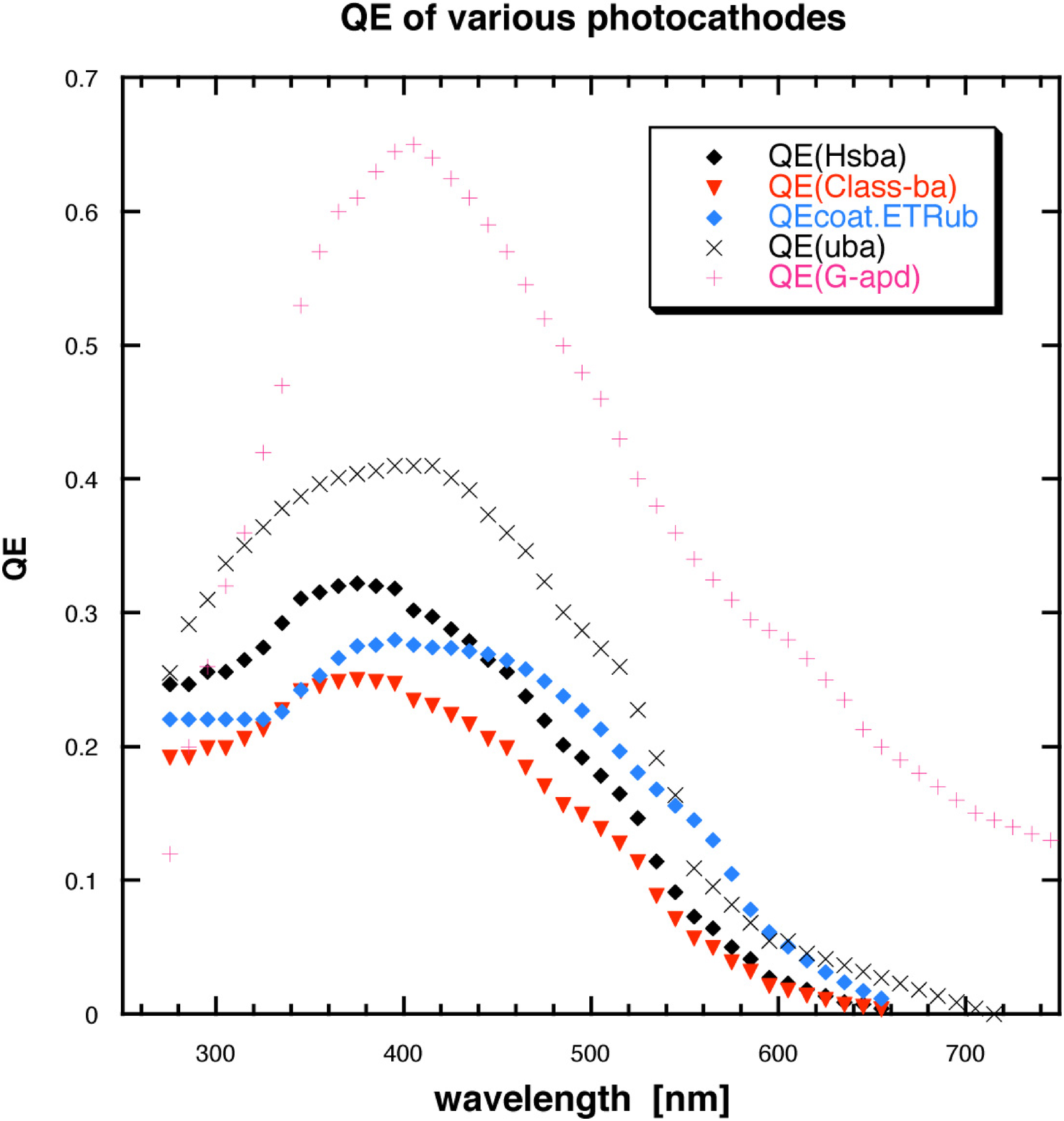}
  \caption{QE ($\lambda$) for different sensors}
  \label{fig2}
 \end{figure}

In order to highlight the possible improvement we compare in Fig. 2 the QE 
($\lambda )$ of a G-APD (3x3 mm Hamamatsu MPPC) with that of some PMTs of 
different cathodes: a) a flat window PMT with a standard bialkali (Ba) 
photocathode, b) the Electron Tube hemispherical PMT 9116 B with a RbCs 
cathode and a special surface treatment (used in the MAGIC I telescope), c) 
a hemispherical Hamamatsu PMT with the novel superbialkali (Sba) cathode and 
d) a flat window PMT with the high QE Ultrabialkali (Uba) cathode.

\begin{figure}[!t]
\centering
  \includegraphics[width=2.1in]{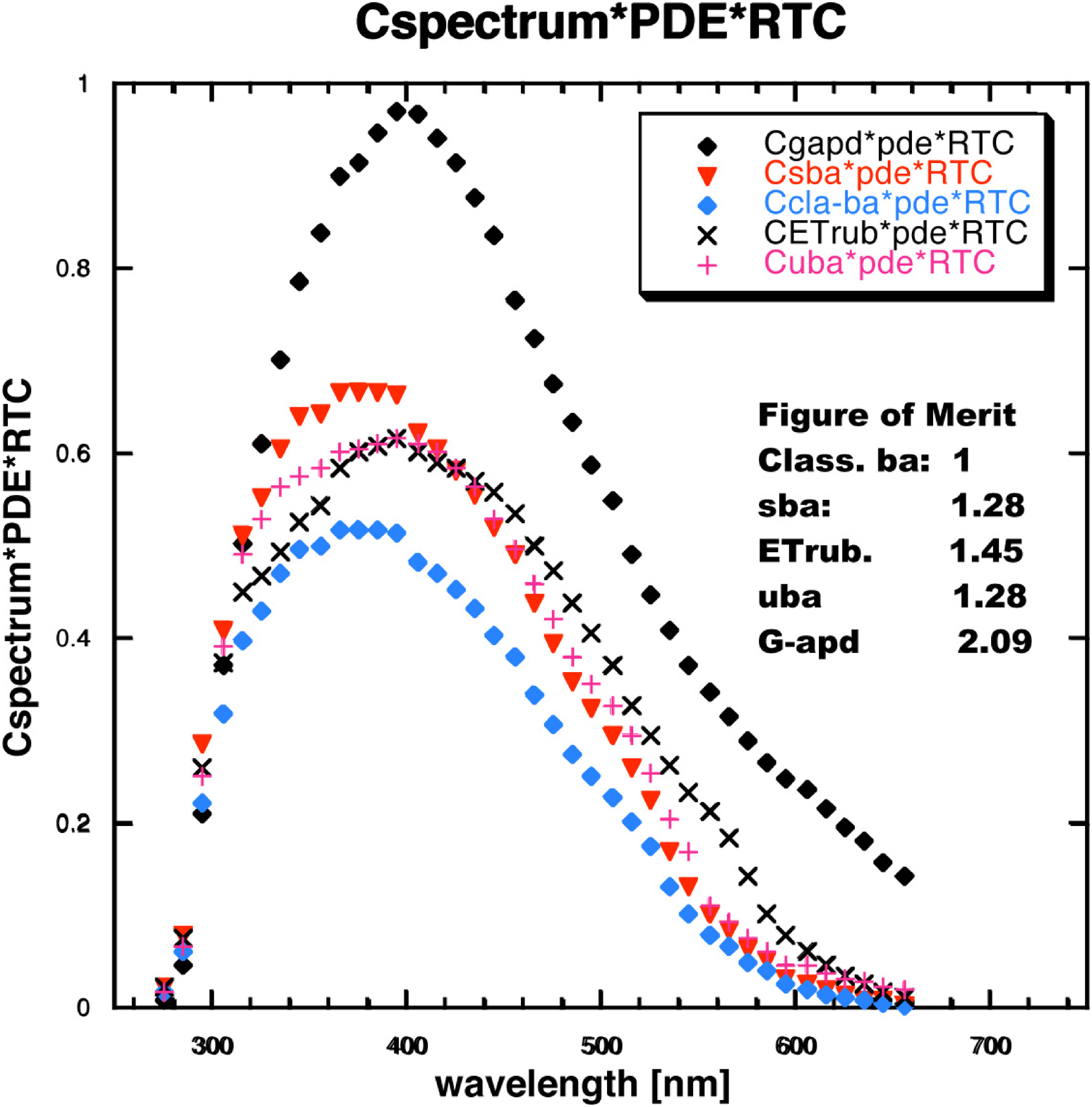}
  \caption{PDE ($\lambda$) folded withthe Cherenkov spectrum and the optical
  parameters (R,T,C) of the Magic telescope}
  \label{fig3}
 \end{figure}

The G-APD could only be operated at 1.3 V V$_{bd}$ because optical crosstalk 
becomes excessive at higher V$_{bd}$. Therefore, the ratio $\varepsilon $ = 
PDE/QE of the G-APD is only 65 {\%}. The flat window standard PMT and the 
hemispherical Sba PMT have an $\varepsilon $ of 0.9, while the RbCs PMT used 
in Magic has an $\varepsilon {\rm g}$ 0.95 (due to a maximized voltage 
between photocathode and 1$^{st}$ dynode) and the flat window Hamamatsu Uba 
PMT with mesh dynodes an $\varepsilon $ of 0.65 \cite{ham}.

 \begin{figure*}[!t]
\centering
  \includegraphics[width=0.8\linewidth]{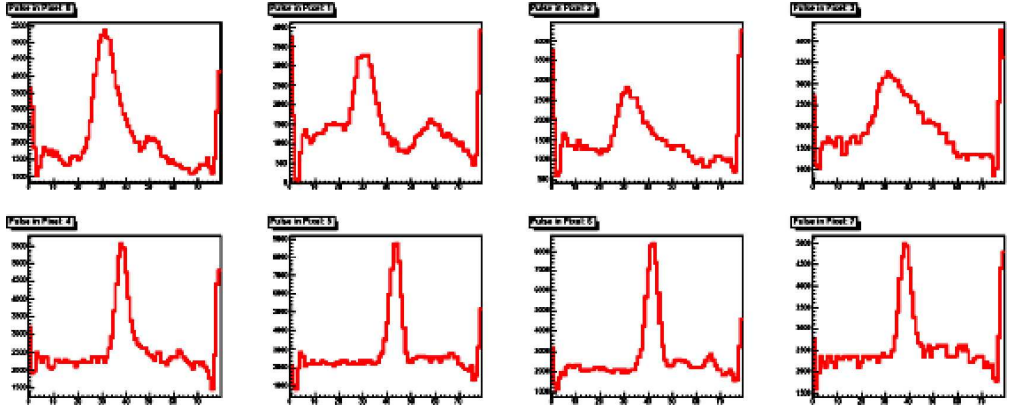}
  \caption{An event seen by the 4 G-APDs (top) and in the surrounding 4 PMTs
  (bottom row), digitized by a 2 Ghz F-Adc. The G-APD signals are clipped by 50
  Ohm coax cables to $\approx 3$ nsec. Pickup peaks at start and end by
  multiplexer.}
  \label{fig4}
 \end{figure*}
 
Fig 3 shows the PDE ($\lambda )$ folded by the Cherenkov spectrum (50 GeV 
$\gamma $ showers at 30\r{ } zenith angle at 2200 m altitude) and the 
reflectivity R of the MAGIC mirror, the camera window transmission W and the 
reflectivity of the light catcher C. We also calculated the so-called figure 
of merit (FM) by integrating PDF($\lambda )$ between 290 and 650 nm and 
normalizing the numbers to the value of the standard Ba flat-window PMT (see 
insert in figure 3). $\varepsilon {\rm g}$s only a guide-number as both 
the G-APDs and PMTs have fluctuations in design parameters and are often 
operated at different voltage configurations.

\section{Some observations }

In order to verify the potential of G-APDs to detect Cherenkov light we have 
carried out a number of observations, \cite{biland}. We used two types of G-APDs, 
the n-on-p type from Photonique (SSPM{\_}0606BG4MM, area 4.4 mm$^{2}$ each, 
peak QE at 580 nm) and the p-on-n Hamamatsu MPPC with 55-65{\%} peak QE at 
450 nm, $\approx $ 70 V bias, 50x50 or 100x100 $\mu $ cell size, $\approx $ 
250 kHz noise/1 mm$^{2}$ area at 26\r{ }C, $\tau _{rise} \quad \approx $ 2 
nsec, $\tau _{fall} \quad \approx $ 30 nsec, see the Hamamatsu data sheet for 
the MPPCs. In all four tests we could detect Cherenkov light with about the 
predicted efficiency. In the fourth test, installing 4 MPPCs enlarged by 6x6 
mm$^{2}$ light catchers in the MAGIC camera and comparing the recorded 
signal with that of the surrounding PMTs, we could confirm an increase in FM 
by about 1.8. Fig. 4 shows a recording of a shower both in the different 
G-APD and PMT pixels. Fig. 5 shows the shower image in the MAGIC camera and 
the location of the test pixels. Fig. 6 shows a larger statistics 
correlation between the PMT and G-APD signals when normalizing to the same 
sensor area. G-APDs detect 1.8 times more PEs/unit area (after optical 
cross-talk correction). In order to highlight the calibration prospects, we 
show in Fig. 7 the G-APD spectra with clear peaks for 1 and 2 (3) PEs.

\section{Some comments related to the deficiencies}

As noted, current G-APDs have some deficiencies, which need to be corrected 
to efficiently use G-APDs in IACT cameras. Here we will discuss only the 
most critical issues. As mentioned, G-APDs have optical crosstalk firing 
neighboring cells when a cell detects a photon or a noise trigger occurs. 
For a gain of 10$^{5}$ typically 3-4 photons of 1-1.5 eV energy are 
generated in an avalanche breakdown. Currently, the G-APD gain can easily 
exceed 10$^{6}$ when the cell area is large and V$_{bd}$ is set well above 1 
V to reach a high PDE, see eq. (1). A prerequisite for a high PDE is to use 
G-APDs with large cells and little dead space and to operate them at high 
overvoltage. This is often impossible because the large photon production at 
high gain resulting in considerable cross-talk and, in turn, in a large 
excess noise. One solution is to lower the cell capacitance by constructive 
means, e.g. by increasing the p-n cell thickness. This allows one to 
increase the maximally allowed overvoltage, i.e. $\varepsilon $ approaches 1 
and (the PDE nearly the QE). As caveat the dark counts will increase but 
also V$_{bd}$ has likely to be increased. Industry is working on such 
changes. Hopefully, a higher FM, close to 3, will be achieved soon. 
Alternatively, one can insert grooves between cells but this normally 
increases the dead area between cells and in turn lowers the average QE. The 
seemingly high noise of even the best G-APDs is not so critical in the case 
of use in large IACT. The night sky light background rate will exceed easily 
2-4 Mhz/mm$^{2}$ G-APD area when observing a dark sky area outside the 
galactic plane. Very noisy G-APDs can be cooled but water condensation has 
to be avoided. Another critical issue is the temperature and voltage 
dependence of the PDE at small overvoltages. The breakdown voltage of the 
Hamamatsu MPPC rises by 50 mV/degree. At 1-2 V overvoltage an increase of a 
few degrees will significantly lower the overvoltage and in turn the PDE. 
Similar changes occur when the bias voltage is not well stabilized. The 
solution is to either stabilize the temperature or correct the bias voltage 
accordingly \cite{miy}. In case the G-APD can be operated at high overvoltage the 
influence of the temperature and voltage drift becomes less critical as the 
G-APD operates already at nearly a plateau in $\varepsilon $. The generally 
low UV sensitivity can be increased by a transparent lacquer coating doped 
with a wavelength shifter. Such improvements have been achieved with PMTs 
\cite{eigen} and also with silicon PIN photodiodes.

 \begin{figure}[!t]
\centering
  \includegraphics[width=2.1in]{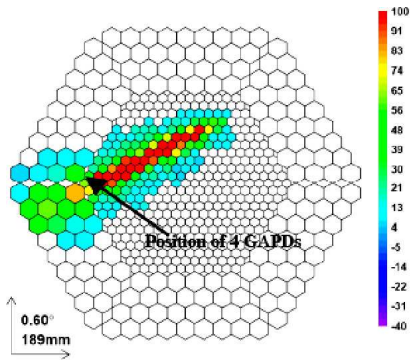}
  \caption{Event recorded by MAGIC, as shown in Fig. 4}
  \label{fig5}
 \end{figure}
 
 \begin{figure}[!t]
\centering
  \includegraphics[width=2in]{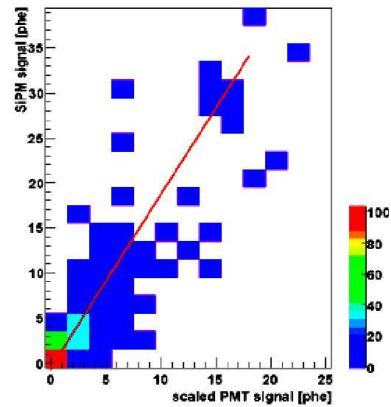}
  \caption{Amplitude correlations of G-APD and PMT signals (300 events),
  normalized to same area}
  \label{fig6}
 \end{figure}

 \begin{figure}[!t]
\centering
  \includegraphics[width=2in]{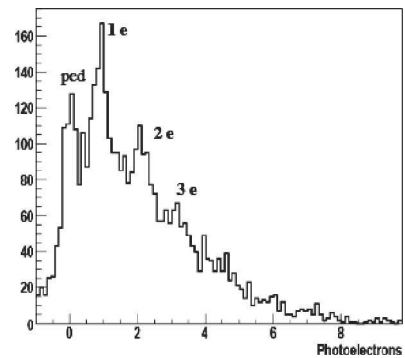}
  \caption{Pulse height spectra of the G-APDs, recorded by thefull MAGIC readout
  chain.}
  \label{fig7}
 \end{figure}

\section{Conclusions and outlook}

The current studies have shown that G-APDs have the potential to become a 
superior photodetector for IACT cameras. The gain of the tested G-APDs is 
too high allowing a large overvoltage to reach a high PDE. The problem is 
linked to high optical crosstalk. If industry succeeds in to lowering the 
cell capacitance at the same cell area one should reach a FM of up to 3 
compared to PMTs with standard Ba cathodes. Also, a high regulation of the 
temperature or of the bias voltage is required.

\end{document}